\documentclass[twocolumn,showpacs,preprintnumbers,amsmath,amssymb]{revtex4}
\usepackage{graphicx}
\usepackage{dcolumn}
\usepackage{bm}
\begin{document}

\title{NEW SYMMETRY IN NUCLEOTIDE SEQUENCES}

\author{Marina A.Makarova}
\author{Michael G.Sadovsky}
\email{msad@icm.krasn.ru}
\affiliation{Institute of biophysics of SD of RAS;\\ 660036
Russia, Krasnoyarsk, Akademgorodok.}

\begin{abstract}
Information valuable words are the strings with the significant deviation of real
frequency from the expected one. The expected frequency is determined through the maximum
entropy principle of the reconstructed (extended) frequency dictionary of strings
composed from the shorter words. The information valuable words are found to be the
complementary palindromes: they are read equally in opposite directions, if nucleotides
are changed for the complementary ones ($\{\mathsf{A}~\rightleftharpoons~\mathsf{T}; \;
\mathsf{C}~\rightleftharpoons~\mathsf{G}\}$) in one of them. Some properties of such
symmetric words are discussed.
\end{abstract}


\maketitle

\section{\label{introd}Introduction}
A study of statistical and informational features of nucleotide sequences provides a
researcher with powerful tool for investigations, and reveals some new issues in
fundamental problems of molecular biology, genetics and evolution. Still, new patterns
could be found in the genetic entities. Here we present such new pattern exhibiting quite
unusual symmetry of the genomes.

Further, we shall consider the continuous sequences, only. A genetic entity is a symbol
sequence from the four-letter alphabet $\aleph = \{\mathsf{A}, \mathsf{C}, \mathsf{G},
\mathsf{T}\}$ with $\mathsf{A}$ referring to adenine, $\mathsf{C}$ referring to cytosine,
$\mathsf{G}$ referring to guanine, and $\mathsf{T}$ referring to thymine. A study of
unbound sequences is possible, while it brings no essential knowledge, but the technical
difficulties.

Consider a nucleotide sequence of the length $N$. Any continuous subsequence $\omega$,
$\omega = \nu_1\nu_2\nu_2\ldots \nu_{q-1}\nu_q$ of the length $q$, $1 \leq q \leq N$
makes a word; here $\nu_j \in \aleph$ is a nucleotide occupying the $j$--th position in
it. Further, we shall consider the words of the length $3 \leq q \leq 8$. A set of the
(different) words of the length $q$ occurred at a sequence is the support of that latter.
Providing each element of the support (i.e., each word $\omega$) with its frequency
$$f_{\omega} = \frac{n_{\omega}}{N}\;,$$ one gets the frequency dictionary
$W_q$ of the sequence; here $n_{\omega}$ is the number of copies of the word $\omega$.
Evidently, an information contained in a frequency dictionary $W_l$ is entirely contained
at the frequency dictionary $W_q$, as $q > l$.

An idea to seek for the overrepresented, or, on the contrary, underrepresented strings
within a nucleotide sequence is not so new \cite{ol1,ol2,ol3,ol4}. The problem arises as
one faces the problem of what is an overrepresentation (or the underrepresentation, in
turn) of a string.

A word $\widetilde{\omega}$ was considered to be an overrepresented one, if the number of
its copies just exceeded some given number $m_{\widetilde{\omega}}$. Such approach fails
to take into account a structure of a sequence. Indeed, if a sequence is rather
degenerated, with the greatest majority of (possible) words from the given alphabet
exhibiting a zero frequency, then few words show a great abundance within a sequence,
while this abundance seems to be quite natural; one has no way to figure out a word of
the increased frequency.

\subsection{Expected frequency of a word}\label{vs}
The method for the estimation of a word frequency taking into account the structure of a
sequence was proposed in \cite{n1,n2,n3}; later, it has been reproduced in \cite{kitai}.
The method estimates a word frequency $\widetilde{f}_{\omega}$ through the calculation of
the most probable continuation of a shorter word $\omega'$ embedded into the given one. A
word $\omega$ of the length $q$ could be composed from a couple of words $\omega'$ and
$\omega''$ of the length $q-1$: $\omega = \omega' \cup \omega''$, so that $\omega' \cap
\omega''$ yields the word $\dot{\omega}$ of the length $q-2$ occupying the central part
of $\omega$.
\begin{table*}
\caption{\label{T3}Information value of triplets of some bacterial genomes.}
\begin{ruledtabular}
\begin{tabular}{c|cccccccccccc}
$\omega$ & 1 & 2 & 3 & 4 & 5 & 6 & 7 & 8 & 9 & 10 & 11 & 12\\
\hline
AAA & 1.24839 & 0.92083 & 1.25386 & 1.01475 & 1.06220 & 1.21886 & 1.08208 & 1.03892 & 1.03863 & 1.03191 & 0.96032 & 1.03133\\
AAC & 0.88993 & 1.12276 & 0.83773 & 0.99549 & 1.08379 & 0.91354 & 0.99041 & 1.01165 & 1.04143 & 0.98565 & 1.11912 & 1.00490\\
AAG & 0.98376 & 1.03584 & 1.01543 & 1.04766 & 0.99776 & 0.98089 & 1.02092 & 1.06553 & 1.03670 & 1.06628 & 0.97427 & 1.12155\\
AAT & 0.92113 & 0.82808 & 0.84938 & 0.95822 & 0.87009 & 0.91847 & 0.88116 & 0.88997 & 0.89139 & 0.91506 & 1.01950 & 0.86986\\
ACA & 0.78792 & 0.87957 & 0.97284 & 1.08486 & 1.19674 & 0.80679 & 1.13178 & 0.80215 & 0.84119 & 0.87144 & 0.88932 & 1.02568\\
ACC & 1.25357 & 1.35953 & 1.09049 & 1.02062 & 0.94054 & 1.26175 & 0.93880 & 1.34158 & 1.26521 & 1.18203 & 1.24732 & 1.00980\\
ACG & 0.96791 & 0.83647 & 1.06375 & 1.03756 & 1.14600 & 0.96004 & 1.23572 & 0.92220 & 0.92207 & 1.18390 & 1.20931 & 1.15087\\
ACT & 0.85948 & 0.87767 & 0.85883 & 0.90253 & 0.80537 & 0.85209 & 0.83204 & 1.03680 & 1.05449 & 0.93759 & 0.97212 & 0.82105\\
AGA & 0.87567 & 0.85369 & 1.15615 & 1.09460 & 1.07654 & 0.86428 & 1.06904 & 0.96863 & 0.95431 & 0.99529 & 0.90745 & 0.90254\\
AGC & 1.04776 & 1.18494 & 0.88060 & 1.03884 & 1.07283 & 1.04863 & 1.02615 & 1.09375 & 1.09344 & 1.09717 & 1.21578 & 1.25251\\
AGG & 1.08438 & 0.94306 & 1.15786 & 0.90758 & 1.00506 & 1.06754 & 1.04211 & 0.84210 & 0.84145 & 0.97922 & 0.99877 & 1.06538\\
AGT & 0.85408 & 0.88046 & 0.85697 & 0.89796 & 0.80160 & 0.89994 & 0.83503 & 1.05057 & 1.06709 & 0.94404 & 0.98228 & 0.81662\\
ATA & 1.06218 & 0.98464 & 1.25295 & 1.10187 & 1.08656 & 1.09104 & 1.09551 & 1.00897 & 0.99696 & 1.13538 & 0.88815 & 1.00338\\
ATC & 1.02810 & 1.24158 & 1.08574 & 1.03599 & 1.06903 & 1.04250 & 1.04908 & 1.28184 & 1.25251 & 1.07035 & 1.23619 & 1.06034\\
ATG & 1.00536 & 0.84958 & 0.91335 & 0.89960 & 1.01971 & 0.94402 & 1.01273 & 0.89525 & 0.91688 & 0.93045 & 0.95151 & 1.11880\\
ATT & 0.91949 & 0.84086 & 0.85922 & 0.95816 & 0.88275 & 0.99782 & 0.89472 & 0.88430 & 0.90041 & 0.91549 & 1.02917 & 0.87751\\
CAA & 0.77210 & 1.07943 & 0.85591 & 1.10052 & 0.88275 & 0.82360 & 0.98988 & 1.02671 & 1.04806 & 1.10549 & 1.03027 & 0.98016\\
CAC & 0.96551 & 0.93073 & 1.15442 & 0.94802 & 0.94057 & 0.93362 & 1.01832 & 1.04749 & 1.04077 & 0.86938 & 1.12423 & 0.95280\\
CAG & 1.16146 & 1.16196 & 1.08065 & 0.94118 & 1.01340 & 1.18275 & 0.97340 & 1.04852 & 1.02862 & 1.03538 & 0.96776 & 0.91271\\
CAT & 1.02809 & 0.84712 & 0.91230 & 0.90690 & 1.01397 & 0.98505 & 1.02669 & 0.88986 & 0.88692 & 0.91658 & 0.93155 & 1.11899\\
CCA & 1.03399 & 1.15939 & 0.87031 & 1.03525 & 1.03783 & 1.01123 & 1.00765 & 1.24131 & 1.18888 & 1.22535 & 1.10523 & 1.03135\\
CCC & 0.93101 & 0.73478 & 1.11178 & 1.09415 & 1.00459 & 0.90676 & 1.01826 & 0.82663 & 0.88008 & 0.75666 & 0.88551 & 0.92618\\
CCG & 1.00599 & 1.14085 & 0.90936 & 1.08087 & 0.92568 & 1.02537 & 0.84738 & 0.99767 & 0.98756 & 0.85005 & 0.65519 & 0.95719\\
CCT & 1.07066 & 0.93317 & 1.15751 & 0.90651 & 1.00145 & 1.09341 & 1.04670 & 0.84436 & 0.87002 & 0.98851 & 1.02769 & 1.06686\\
CGA & 1.12216 & 1.19532 & 0.77033 & 0.86699 & 0.94725 & 1.12191 & 0.94181 & 1.01782 & 1.01351 & 0.95671 & 0.94004 & 1.03074\\
CGC & 0.95479 & 0.86849 & 1.17801 & 1.06972 & 1.01305 & 0.95573 & 0.96489 & 1.04952 & 1.05949 & 0.98305 & 0.99573 & 0.85999\\
CGG & 1.00656 & 1.13720 & 0.90336 & 1.09972 & 0.89048 & 1.00923 & 0.86318 & 1.00714 & 0.99760 & 0.84705 & 0.65418 & 0.93832\\
CGT & 0.97038 & 0.83977 & 1.07212 & 1.05974 & 1.17105 & 0.95601 & 1.24474 & 0.90951 & 0.91183 & 1.19404 & 1.21322 & 1.15658\\
CTA & 0.78076 & 1.06138 & 0.72074 & 0.90150 & 0.97012 & 0.80087 & 0.96955 & 0.83784 & 0.87414 & 0.99675 & 1.11544 & 0.91193\\
CTC & 0.88475 & 0.80308 & 1.09278 & 1.11410 & 1.02273 & 0.88729 & 1.01232 & 0.97523 & 1.03281 & 0.86614 & 0.84845 & 0.98217\\
CTG & 1.16691 & 1.16746 & 1.07107 & 0.93944 & 1.00497 & 1.18212 & 0.99626 & 1.04533 & 1.02536 & 1.03842 & 0.99525 & 0.91585\\
CTT & 0.97196 & 1.01318 & 1.01568 & 1.04373 & 1.00103 & 0.92393 & 1.01426 & 1.07071 & 1.02737 & 1.06123 & 0.96421 & 1.11788\\
GAA & 1.16532 & 1.04307 & 1.05977 & 0.94460 & 1.03518 & 1.12957 & 1.01435 & 0.91193 & 0.90301 & 0.99779 & 0.95390 & 1.00629\\
GAC & 1.00689 & 0.97547 & 0.76977 & 0.93615 & 0.77792 & 1.02613 & 0.86176 & 0.80629 & 0.79984 & 1.05900 & 0.72483 & 0.91725\\
GAG & 0.88854 & 0.79737 & 1.09270 & 1.11447 & 1.01863 & 0.89367 & 1.00374 & 0.96934 & 1.02206 & 0.86479 & 0.81669 & 0.98885\\
GAT & 1.00542 & 1.24808 & 1.08388 & 1.04077 & 1.07054 & 0.99739 & 1.05915 & 1.28507 & 1.25373 & 1.07314 & 1.20473 & 1.05493\\
GCA & 0.93875 & 0.92025 & 1.15216 & 0.98651 & 1.07677 & 0.95435 & 1.11102 & 0.86063 & 0.84900 & 0.91414 & 0.91229 & 0.91695\\
GCC & 1.06046 & 1.14680 & 0.71514 & 0.93333 & 0.78368 & 1.07192 & 0.80541 & 1.03673 & 1.03993 & 0.98342 & 0.75770 & 0.96929\\
GCG & 0.95537 & 0.86589 & 1.17244 & 1.05271 & 1.00356 & 0.94149 & 0.98456 & 1.04628 & 1.05970 & 0.97297 & 1.01888 & 0.85792\\
GCT & 1.05902 & 1.18556 & 0.88587 & 1.03560 & 1.06550 & 1.05257 & 1.03468 & 1.09813 & 1.10238 & 1.10919 & 1.22074 & 1.25342\\
GGA & 0.74898 & 0.78464 & 1.20166 & 0.99219 & 1.19549 & 0.78189 & 1.11571 & 0.79437 & 0.80350 & 1.02460 & 0.93277 & 1.06861\\
GGC & 1.06266 & 1.14162 & 0.71231 & 0.91135 & 0.78061 & 1.07411 & 0.82776 & 1.03906 & 1.04125 & 0.98142 & 0.71547 & 0.96591\\
GGG & 0.92302 & 0.73480 & 1.11195 & 1.09543 & 1.01156 & 0.90622 & 1.04257 & 0.82446 & 0.84210 & 0.75753 & 0.87182 & 0.93187\\
GGT & 1.25708 & 1.36092 & 1.09351 & 1.01592 & 0.92056 & 1.24171 & 0.95112 & 1.32290 & 1.29006 & 1.18076 & 1.28108 & 1.00632\\
GTA & 1.37813 & 1.12428 & 1.30753 & 1.05004 & 1.17874 & 1.25460 & 1.14970 & 1.15108 & 1.09363 & 1.08213 & 0.87842 & 1.13185\\
GTC & 0.99903 & 0.98226 & 0.77452 & 0.93243 & 0.79476 & 0.98748 & 0.85875 & 0.80133 & 0.80606 & 1.07499 & 0.71102 & 0.92286\\
GTG & 0.97806 & 0.91586 & 1.14528 & 0.94913 & 0.95182 & 0.99491 & 0.98758 & 1.04272 & 1.02745 & 0.86433 & 1.12254 & 0.97126\\
GTT & 0.89238 & 1.13658 & 0.85019 & 1.00812 & 1.05307 & 0.92245 & 1.00269 & 1.01886 & 1.05596 & 0.98775 & 1.11781 & 0.99816\\
TAA & 0.73264 & 0.69552 & 0.75517 & 0.95203 & 0.86201 & 0.69853 & 0.87456 & 1.00342 & 0.97725 & 0.84490 & 1.06262 & 0.95946\\
TAC & 1.41080 & 1.11920 & 1.32854 & 1.06826 & 1.16044 & 1.40841 & 1.13754 & 1.15015 & 1.10730 & 1.09494 & 0.87586 & 1.15052\\
TAG & 0.75912 & 1.06564 & 0.70612 & 0.89680 & 0.97181 & 0.67018 & 0.98811 & 0.85319 & 0.86640 & 0.99320 & 1.13028 & 0.89961\\
TAT & 1.04889 & 0.98748 & 1.26476 & 1.09409 & 1.11020 & 1.28451 & 1.09117 & 0.99799 & 1.05394 & 1.14742 & 0.92050 & 1.02361\\
TCA & 1.27939 & 1.02063 & 0.93763 & 0.93237 & 0.79309 & 1.28009 & 0.83897 & 1.15355 & 1.17408 & 1.01336 & 1.11127 & 1.02233\\
TCC & 0.74532 & 0.78513 & 1.20897 & 0.97646 & 1.18581 & 0.73439 & 1.15114 & 0.78603 & 0.80854 & 1.02128 & 0.93670 & 1.06748\\
TCG & 1.12337 & 1.19362 & 0.77605 & 0.90285 & 0.94649 & 1.13474 & 0.94982 & 1.01570 & 1.00775 & 0.96797 & 0.92383 & 1.02323\\
TCT & 0.86843 & 0.87130 & 1.15036 & 1.09314 & 1.08316 & 0.85471 & 1.05942 & 0.97370 & 0.93210 & 0.98650 & 0.89900 & 0.89596\\
TGA & 1.26658 & 1.02426 & 0.94177 & 0.93123 & 0.79009 & 1.22408 & 0.85859 & 1.15023 & 1.15598 & 1.00702 & 1.10190 & 1.00949\\
TGC & 0.94732 & 0.92705 & 1.15522 & 0.99118 & 1.06657 & 0.92706 & 1.11411 & 0.86068 & 0.85578 & 0.92404 & 0.93901 & 0.91540\\
TGG & 1.03319 & 1.15125 & 0.87639 & 1.02891 & 1.04896 & 1.06676 & 0.98687 & 1.23625 & 1.22838 & 1.23514 & 1.12880 & 1.04403\\
TGT & 0.78350 & 0.88090 & 0.96292 & 1.08616 & 1.20455 & 0.80066 & 1.11386 & 0.81371 & 0.82204 & 0.86127 & 0.86943 & 1.03059\\
TTA & 0.73902 & 0.69607 & 0.76506 & 0.95066 & 0.87277 & 0.88349 & 0.87928 & 1.00069 & 1.02199 & 0.85664 & 1.09307 & 0.97715\\
TTC & 1.16938 & 1.02814 & 1.05823 & 0.94957 & 1.02477 & 1.17741 & 1.01877 & 0.91987 & 0.89988 & 0.99132 & 0.92495 & 1.00273\\
TTG & 0.75735 & 1.09836 & 0.87109 & 1.10405 & 1.00338 & 0.74185 & 0.99859 & 1.02627 & 1.03366 & 1.09651 & 1.00745 & 0.96749\\
TTT & 1.26647 & 0.91158 & 1.23506 & 1.01272 & 1.06513 & 1.24765 & 1.06942 & 1.03415 & 1.02698 & 1.03396 & 0.95441 & 1.03250\\
\end{tabular}
\end{ruledtabular}
\end{table*}

Such combinations must meet the following linear constraints: combination $\omega =
\omega' \cup \omega''$ must produce the frequency dictionary $\widehat{W}_q$, that yields
the real frequency dictionary $W_{q-1}$ upon the downward conversion. This linear
constraint eliminates a part of possible combinations, but the diversity of these latter
is still great enough. The diversity of combinations results from an ambiguity of
continuations of a shorter word met within a sequence, in general. Thus, one has to
figure out a single frequency dictionary $\widetilde{W}_q$ from the family
$\{\widetilde{W}_q\}$. It seems rather natural to take into consideration the dictionary
$\widetilde{W}_q$ bearing the most probable continuations of the words of the length
$q-1$. The entity is identified due to entropy $S = - \sum_{\omega}
\widetilde{f}_{\omega} \cdot \ln \widetilde{f}_{\omega}$; the dictionary bearing the most
probable continuations has the maximal entropy within the family $\{\widetilde{W}_q\}$.

This extremal principle provides a researcher with the explicit formula for the expected
frequency:
\begin{subequations}\label{vosfr}
\begin{equation}\label{vosf-q}
\widetilde{f}_{\nu_1\nu_2\nu_3\ldots \nu_{q-1}\nu_q} = \frac{f_{\nu_1\nu_2\ldots
\nu_{q-2}\nu_{q-1}} \times f_{\nu_2\nu_3\ldots \nu_{q-1}\nu_q}}{f_{\nu_2\nu_3\ldots
\nu_{q-2}\nu_{q-1}}}\;,
\end{equation}
\textrm{or}
\begin{equation}\label{vosf2}
\widetilde{f}_{\nu_1\nu_2} = f_{\nu_1} \times f_{\nu_2}
\end{equation}
\end{subequations}
for the case of $q=2$. Formula~(\ref{vosf-q}) seems to be similar to the formula of
Markov process transition probability. Nevertheless, the formulae~(\ref{vosfr}) are
developed for a sequence, which is not hypothesized for being a random Markov process.
Formulae~(\ref{vosfr}) mean that the Markov process (of the corresponding order) realizes
the hypothesis of the most probable continuation of a word. The formulae~(\ref{vosfr})
could be generalized for the case of reconstruction of the dictionary $\widetilde{W}_q$
over the original dictionary $W_l$, with $l < q-1$, while we shall not consider that
case. Everywhere below the reconstruction of $\widetilde{W}_q$ over $W_{q-1}$.

\subsection{Information valuable words}\label{inkap} The information valuable words
are the strings with significant deviation of the real frequency from the expected one.
Such strings contribute an information capacity of a frequency dictionary most of all
\cite{s1-1,s1-2}. The information capacity is defined as a mutual entropy of the given
dictionary $W_q$ calculated with respect to the reconstructed one $\widetilde{W}_q$. Such
definition holds true, since the reconstructed frequency dictionary $\widetilde{W}_q$
bears the same words, as the real one $W_q$ does, and, maybe, some strings else.

The informal definition should be replaced with more strict and rigorous: a string
$\omega^{\ast}$ is the information valuable word, if it falls out of the range determined
by the double inequality
\begin{equation}\label{alf}
\underline{\alpha} \leq \frac{f_{\omega^{\ast}}}{\widetilde{f}_{\omega^{\ast}}} \leq
\overline{\alpha}\; .
\end{equation}
Here $0< \underline{\alpha} < 1$ and $\overline{\alpha} > 1$ are the upper and lower
information value threshold, respectively. The motivation behind such definition of
information valuable word is clear and obvious. An expected frequency
$\widetilde{f}_{\omega}$ determined according to (\ref{vosfr}) meets the following
constraint: $\widetilde{f}_{\omega} > 0$, if $f_{\omega}>0$, for any $\omega$. Thus, the
definition (\ref{alf}) is correct.

Whether a word $\omega$ is of information value, or not, depends strongly on the choice
of the threshold. In general, the choice is to be done with respect to a variety of
issues, including the specific targets of a study. Thus, the choice of $\alpha$ could be
done in various ways, and there is no single natural method to identify that latter. The
problem could be addressed studying a distribution of the words over their information
value $p_{\omega} = f_{\omega}/\widetilde{f}_{\omega}$; Table~\ref{T3} shows such
distribution for $q=3$, for a dozen of bacterial genomes (all the entities are deposited
at EMBL--bank, ({\em http://www.ebi.ac.uk/genomes}): 1 -- {\sl Deinococcus radiodurans},
chromosome I (identifier AE000513); 2 -- {\sl Mycobacterium tuberculosis CDC1551}
(identifier AE000516); 3 -- {\sl Treponema pallidum} (identifier AE000520); 4 -- {\sl
Borrelia burgdorferi B31} (identifier AE000783); 5 -- {\sl Chlamydia trachomatis
D/UW-3/CX} (identifier AE001273); 6 -- {\sl Deinococcus radiodurans}, chromosome II
(identifier AE001825); 7 -- {\sl Chlamydophila pneumoniae AR39} (identifier AE002161); 8
-- {\sl Vibrio cholerae O1}, chromosome I (identifier AE003852); 9 -- {\sl Vibrio
cholerae O1}, chromosome II (identifier AE003853); 10 -- {\sl Streptococcus pneumoniae
R6} (identifier AE007317); 11 -- {\sl Ureaplasma parvum} (identifier AF222894); and 12 --
{\sl Bacillus str. 168} (identifier AL009126).
\begin{table}
\caption{\label{T1}Complementary palindromic triplets of six eukaryotic genomes.}
\begin{ruledtabular}
\begin{tabular}{cc|cc}
$f > \alpha \cdot \widetilde{f}$ & $f < \alpha^{-1} \cdot \widetilde{f}$ & $f > \alpha
\cdot \widetilde{f}$ & $f < \alpha^{-1} \cdot \widetilde{f}$
\\ \hline
\multicolumn{2}{c|}{{\sl Debaryomyces hansenii}} & \multicolumn{2}{c}{{\sl
Encephalitozoon cuniculi}} \\ $\mathsf{ACC} \rightleftharpoons \mathsf{GGT}$ &
$\mathsf{CCC} \rightleftharpoons
\mathsf{GGG}$ & $\mathsf{AAG} \rightleftharpoons \mathsf{CCT}$ & $\mathsf{AAT} \rightleftharpoons \mathsf{ATT}$ \\
$\mathsf{TGG} \rightleftharpoons \mathsf{CCA}$ & $\mathsf{CGC} \rightleftharpoons
\mathsf{GCG}$ & $\mathsf{ACA} \rightleftharpoons \mathsf{TGT}$ & $\mathsf{ACT}
\rightleftharpoons \mathsf{AGT}$ \\ \cline{1-2} \multicolumn{2}{c|}{{\sl Candida
glabrata}} & $\mathsf{ATA} \rightleftharpoons \mathsf{TAT}$ & $\mathsf{CTA}
\rightleftharpoons \mathsf{TAG}$ \\ $\mathsf{ATA} \rightleftharpoons \mathsf{TAT}$ &
$\mathsf{TAA} \rightleftharpoons \mathsf{TTA}$ & $\mathsf{GTA} \rightleftharpoons
\mathsf{TAC}$ & $\mathsf{TCA} \rightleftharpoons \mathsf{TGA}$\\ \hline
\multicolumn{2}{c|}{{\sl Eremothecium gossypii}} & \multicolumn{2}{c}{{\sl
Saccharomyces cerevisiae}} \\
$\mathsf{ATA} \rightleftharpoons \mathsf{TAT}$ & $\mathsf{CCC} \rightleftharpoons
\mathsf{GGG}$ & $\mathsf{ATA} \rightleftharpoons \mathsf{TAT}$ & $\mathsf{CCC}
\rightleftharpoons \mathsf{GGG}$ \\ {} & $\mathsf{CTA} \rightleftharpoons \mathsf{TAG}$ &
$\mathsf{CCA} \rightleftharpoons \mathsf{TGG}$ & $\mathsf{TAA} \rightleftharpoons
\mathsf{TTA}$\\ {} & $\mathsf{TAA} \rightleftharpoons \mathsf{TTA}$ & $\mathsf{GTA}
\rightleftharpoons \mathsf{TAC}$ & {} \\ \hline \multicolumn{2}{c|}{{\sl Kluyveromyces
lactis}} & \multicolumn{2}{c}{{\sl Yarrowia lipolytica}} \\
$\mathsf{CCA} \rightleftharpoons \mathsf{TGG}$ & $\mathsf{TAA} \rightleftharpoons
\mathsf{TTA}$ & $\mathsf{ATA} \rightleftharpoons \mathsf{TAT}$ & $\mathsf{CTA}
\rightleftharpoons \mathsf{TAG}$ \\ $\mathsf{GTA} \rightleftharpoons \mathsf{TAC}$ &
$\mathsf{CGC} \rightleftharpoons \mathsf{GCG}$ & $\mathsf{CCA} \rightleftharpoons
\mathsf{TGG}$ & $\mathsf{TAA} \rightleftharpoons \mathsf{TTA}$ \\ $\mathsf{ACC}
\rightleftharpoons \mathsf{GGT}$ & $\mathsf{GAC} \rightleftharpoons \mathsf{GTC}$ &
$\mathsf{GTA} \rightleftharpoons \mathsf{TAC}$ & $\mathsf{TCA} \rightleftharpoons
\mathsf{TGA}$\\ {} & $\mathsf{CCC} \rightleftharpoons \mathsf{GGG}$ & {} & {}\\
\end{tabular}
\end{ruledtabular}
\end{table}

Thicker dictionary could hardly be shown as a table, since the abundance of the words
grows exponentially. Nevertheless, not discussing here the problem of the threshold
choice in detail, further we shall consider the words exhibiting the greatest possible
(within a dictionary $W_q$), or the lowest possible information value $p_{\omega}$,
$p_{\omega} = f_{\omega} / \widetilde{f}_{\omega}$. Surely, the threshold depends on the
dictionary thickness $q$, as well.

\section{Symmetry in genomes}\label{rest}
Genomes of organisms exhibit a symmetry in information valuable words. These latter are
the complementary palindromes. A string, that could be read equally both forward, and
back, is a palindrome. Two words make the complementary palindrome, if the latter is read
equally to the former, upon the complementary replacement of the nucleotides, according
to the complementary rule
$$\{\mathsf{A}~\rightleftharpoons~\mathsf{T}; \;
\mathsf{C}~\rightleftharpoons~\mathsf{G}\}\,.$$ Following is an example of such
complementary palindrome of the length $q=8$:
$$\mathsf{TAGGTAGG}~\rightleftharpoons~\mathsf{CCTACCTA}\;.$$ This is a real complementary
palindrome observed at the {\sl B.subtilis} genome. The information value of the words is
$p_{\mathsf{TAGGTAGG}} = 2.25608$, and $p_{\mathsf{CCTACCTA}} = 2.20500$, respectively.
\begin{table}[t]
\caption{Palindromic triplets of {\sl E.coli}~K-12 complete genome.}\label{eck12-3}
\begin{ruledtabular}
\begin{tabular}{rcl|rcl}
\multicolumn{3}{c|}{$f_{\omega}> \widetilde{f}_{\omega}$} &
\multicolumn{3}{c}{$f_{\omega} < \widetilde{f}_{\omega}$} \\ \hline $p_{left}$ & {} &
$p_{right}$ & $p_{left}$ & {} & $p_{right}$ \\ \hline 1.54765 & $\mathsf{CAG}
\rightleftharpoons \mathsf{CTG}$ & 1.54368 & 0.61717 & $\mathsf{TAG} \rightleftharpoons
\mathsf{CTA}$ & 0.61033\\ 1.26703 & $\mathsf{ACC} \rightleftharpoons \mathsf{GGT}$ &
1.26644 & 0.61717 & $\mathsf{CTC} \rightleftharpoons \mathsf{GAG}$ & 0.61033\\ 1.19401 &
$\mathsf{GAT} \rightleftharpoons \mathsf{ATC}$ & 1.19159 & 0.76580 & $\mathsf{GGG}
\rightleftharpoons \mathsf{CCC}$ & 0.76341 \\ 1.15438 & $\mathsf{CCA} \rightleftharpoons
\mathsf{TGG}$ & 1.15122 & {} & {} & {} \\
\end{tabular}
\end{ruledtabular}
\end{table}

Consider Table~\ref{T3} in more detail. It is evident, that the triplets with the highest
information value $p$ make the couples of complementary palindromes. Indeed, these are
the couples $\mathsf{AAA} \rightleftharpoons \mathsf{TTT}$ for the entities \# 1, 3 and
6. The palindrome $\mathsf{ACC} \rightleftharpoons \mathsf{GGT}$ exhibits the increased
level of $p$ for the entities \# 1, 2, 6, 8, 9, 11; etc. Similar pattern is observed for
the triplets with $p < \alpha^{-1}$: the complementary palindrome $\mathsf{TTA}
\rightleftharpoons \mathsf{TAA}$ is observed the entities 1 through 3, the palindrome
$\mathsf{ACA} \rightleftharpoons \mathsf{TGT}$ is observed in the entities 1 and 2, the
couple of triplets $\mathsf{CAA} \rightleftharpoons \mathsf{TTG}$ makes the pattern under
discussion for the entities 1 and 6, and so on.

For each word $\omega$ at the dictionary $W_q$ the information value $p_{\omega} =
f_{\omega}/\widetilde{f}_{\omega}$ could be determined. An excess of the valued over the
information threshold $\alpha$ makes a word to an information valuable one. Since the
determination of the information valuable words depends on the threshold $\alpha$,
further we shall omit this problem through the consideration of the ultimate valued of
$p_{\omega}$, only. The words within a dictionary could be arranged in descending order,
and the words with the highest (or the lowest one, in turn) possible value of
$p_{\omega}$ would be considered. Table~\ref{T1} shows the trinucleotide complementary
palindromes observed in several eukaryotic genomes. The pattern similar to that one shown
in Table~\ref{T1} could be observed in any genome.

Longer words exhibit the same pattern in a duality, as their information value exceeds
the threshold. We shall consider {\sl Escherichia coli}~K-12 complete genome to see the
issue. To begin with, let us consider the complementary palindromic triplets;
Table~\ref{eck12-3} shows the palindromes, and the information value $p =
f_{\omega}/\widetilde{f}_{\omega}$ of the words. The complementary palindromes of the
length $q=4$ and $q=5$ are shown in Table~\ref{eck12-4}. There are two perfect
palindromes of the length $q=4$:
$$\mathsf{CTAG} \quad \textrm{and} \quad \mathsf{GGCC} \; ;$$ with $p_{\mathsf{CTAG}} =
0.25788$ and $p_{\mathsf{GGCC}} = 0.56348$, respectively. A word $\overline{\omega}$ is
the perfect palindrome, if it coincides to itself after the complementary palindromic
transformation.
\begin{table}[h]
\caption{Complementary palindromes of the length $4 \leq q \leq 6$ of {\sl E.coli}~K-12
complete genome.}\label{eck12-4}
\begin{ruledtabular}
\begin{tabular}{rcl}
$p_{left}$ & $f_{\omega}> \widetilde{f}_{\omega}$ & $p_{right}$\\ 1.39567 &
$\mathsf{GTAG} \rightleftharpoons \mathsf{CTAC}$ & 1.37354 \\ 1.33672 & $\mathsf{CTTC}
\rightleftharpoons \mathsf{GAAG}$ & 1.33461 \\ 1.25500 & $\mathsf{CGCC}
\rightleftharpoons \mathsf{GGCG}$ & 1.25360 \\ \hline  {} & $f_{\omega} <
\widetilde{f}_{\omega}$ & {} \\ 0.66683 & $\mathsf{CCAA} \rightleftharpoons
\mathsf{TTGG}$ & 0.66630 \\ 0.67468 & $\mathsf{CAAG} \rightleftharpoons \mathsf{CTTG}$ &
0.66656 \\ \hline {} & $f_{\omega} > \widetilde{f}_{\omega}$ & {} \\ 1.49606 &
$\mathsf{AGGCC} \rightleftharpoons \mathsf{GGCCT}$ & 1.48591 \\ 1.36285 & $\mathsf{TCCGG}
\rightleftharpoons \mathsf{CCGGA}$ & 1.35398 \\ 1.30224 & $\mathsf{GAGGA}
\rightleftharpoons \mathsf{TCCTC}$ & 1.29171 \\ 1.28907 & $\mathsf{GGGGA}
\rightleftharpoons \mathsf{TCCCC}$ &
1.27896 \\ 1.28193 & $\mathsf{GTCTG} \rightleftharpoons \mathsf{CAGAC}$ & 1.28061 \\
1.26112 & $\mathsf{CTCTA} \rightleftharpoons \mathsf{TAGAG}$ & 1.25548 \\ \hline {} &
$f_{\omega} < \widetilde{f}_{\omega}$ & {} \\ 0.50534 & $\mathsf{TTGGA}
\rightleftharpoons \mathsf{TCCAA}$ & 0.47804 \\ 0.60491 & $\mathsf{TCGGA}
\rightleftharpoons \mathsf{TCCGA}$ & 0.61949 \\ 0.60318 & $\mathsf{TAGGA}
\rightleftharpoons \mathsf{TCCTA}$ & 0.65544 \\ \hline {} &
$f_{\omega} > \widetilde{f}_{\omega}$ & {} \\
1.51940 & $\mathsf{TCCGGC} \rightleftharpoons \mathsf{GCCGGA}$ & 1.50514\\
1.48393 & $\mathsf{GGCGCT} \rightleftharpoons \mathsf{AGCGCC}$ & 1.46467\\
1.44442 & $\mathsf{CGGCCC} \rightleftharpoons \mathsf{GGGCCG}$ & 1.43970\\
1.44038 & $\mathsf{TCCTTA} \rightleftharpoons \mathsf{TAACTA}$ & 1.38627\\ \hline {} &
$f_{\omega} < \widetilde{f}_{\omega}$ & {} \\
0.36981 & $\mathsf{GAGACC} \rightleftharpoons \mathsf{GGTCTC}$ & 0.37344 \\ \hline
\end{tabular}
\end{ruledtabular}
\end{table}
Again, it should be stressed, that the complementary palindromes of these lengths are
identified for the ultimate score (upper or lower one) of the information value.

There are a good few perfect complementary palindromes of the length $q=6$. The
hexanucleotides of the ultimate information value (from both sides of the distribution)
are the entities of that type: $$
\begin{array}{lc}
\mathsf{TCTAGA}~(1.54082) & \mathsf{GGCGCC}~(0.04559) \\
\mathsf{GCCGGC}~(0.16533) & \mathsf{GGGCCC}~(0.19223) \\
\mathsf{CACGTG}~(0.27152) & \mathsf{GAGCTC}~(0.30626) \\
\mathsf{CGGCCG}~(0.33055)\;. & {} \\
\end{array}
$$ Surprisingly, the perfect complementary palindromes occupy the leading positions, with
respect to the information value of words observed in this genome at $q=6$.

Complementary palindromes of the length $q=7$ and $q=8$ exhibit more sporadic behavior.
First of all, a lesser part of the information valuable words of that lengths meet a
counter word making a complementary palindrome: some words have no palindromic couple.
Meanwhile, there are 246 complementary palindromes among the information valuable words
of the length $q=7$ with $p > 1$; the abundance of the set of these latter exceeds seven
hundred strings. Similarly, the abundance of the information valuable words with $p < 1$
is more, than nine hundred of strings, while the total number of complementary
palindromes among them is equal to 456 ones. Five highly ranked complementary palindromes
of the length $q=7$ with $p_{\omega} > 1$, and similar palindromes with $p_{\omega} < 1$
are shown in Table~\ref{T-eck12-7}.
\begin{table}[h]
\caption{Complementary palindromes of the length $q=7$ of {\sl E.coli}~K-12 complete
genome.}\label{T-eck12-7}
\begin{ruledtabular}
\begin{tabular}{rcl}
$p_{left}$ & $f_{\omega}> \widetilde{f}_{\omega}$ & $p_{right}$\\ 1.96780 & $\mathsf{GGACTAG} \rightleftharpoons \mathsf{CTAGTCC}$ & 1.64984\\
1.92188 & $\mathsf{CCTAGGG} \rightleftharpoons \mathsf{CCCTAGG}$ & 1.52534\\
1.87437 & $\mathsf{AGACTAG} \rightleftharpoons \mathsf{CTAGTCT}$ & 1.27266\\
1.87146 & $\mathsf{ATTCTAG} \rightleftharpoons \mathsf{CTAGAAT}$ & 1.75984\\
1.86362 & $\mathsf{ATTCCAA} \rightleftharpoons \mathsf{TTGGAAT}$ & 1.60054\\ \hline
\multicolumn{3}{c}{$f_{\omega} < \widetilde{f}_{\omega}$}\\
0.30604 & $\mathsf{GTCTAGG} \rightleftharpoons \mathsf{CCTAGAC}$ & 0.62174\\
0.32277 & $\mathsf{ACCCTAG} \rightleftharpoons \mathsf{CTAGGGT}$ & 0.73860\\
0.32960 & $\mathsf{GGCCTAG} \rightleftharpoons \mathsf{CTAGGCC}$ & 0.48121\\
0.35944 & $\mathsf{CTAGGAA} \rightleftharpoons \mathsf{TTCCTAG}$ & 0.73672\\
0.44693 & $\mathsf{CTACTAG} \rightleftharpoons \mathsf{CTAGTAG}$ & 0.81699\\ \hline
\multicolumn{3}{c}{inversions} \\
1.93827 & $\mathsf{GTCTAGA} \rightleftharpoons \mathsf{TCTAGAC}$ & 0.73333\\
1.92799 & $\mathsf{GCCTAGA} \rightleftharpoons \mathsf{TCTAGGC}$ & 0.77684\\
1.90402 & $\mathsf{TCTAGGA} \rightleftharpoons \mathsf{TCCTAGA}$ & 0.53416\\
1.70238 & $\mathsf{ACTAGAC} \rightleftharpoons \mathsf{GTCTAGT}$ & 0.77531\\
1.65385 & $\mathsf{ACCTAGG} \rightleftharpoons \mathsf{CCTAGGT}$ & 0.27455\\
\end{tabular}
\end{ruledtabular}
\end{table}

Besides, the complementary palindromes of these lengths exhibit an inversion. It means,
that a couple of words may consist of the strings with opposite ration of real and
expected frequency. There are twelve inverse complementary palindromes of the length
$q=7$; Table~\ref{T-eck12-7} shows them, as well.

The complementary palindromes of the length $q=8$ are quite abundant. There is a single
word $\mathsf{TAGGCCTA}$, which is the perfect complementary palindrome; it has the
information value $p = 0.07813$. Table~\ref{T-eck12-8} shows complementary palindromes of
this length, occupying the ultimate positions in the word distribution over the
information value.
\begin{table}[h]
\caption{Complementary palindromes of the length $q=8$ of {\sl E.coli}~K-12 complete
genome.}\label{T-eck12-8}
\begin{ruledtabular}
\begin{tabular}{rcl}
$p_{left}$ & {} & $p_{right}$\\ \hline
\multicolumn{3}{c}{$f_{\omega} > \widetilde{f}_{\omega}$}\\
4.58874 & $\mathsf{CTAGTCTA} \rightleftharpoons \mathsf{TAGACTAG}$ & 3.27632\\
4.36508 & $\mathsf{GCTCCTAG} \rightleftharpoons \mathsf{CTAGGAGC}$ & 3.30159\\
4.00000 & $\mathsf{CCTAGGTG} \rightleftharpoons \mathsf{CACCTAGG}$ & 2.43750\\
3.50000 & $\mathsf{ATCCTAGG} \rightleftharpoons \mathsf{CCTAGGAT}$ & 1.70000\\
3.23214 & $\mathsf{CTAGGAAC} \rightleftharpoons \mathsf{GTTCCTAG}$ & 2.18286\\
3.17143 & $\mathsf{TACTAGGA} \rightleftharpoons \mathsf{TCCTAGTA}$ & 2.11429\\
3.08000 & $\mathsf{CCTAGTAC} \rightleftharpoons \mathsf{GTACTAGG}$ & 2.19048\\
3.04959 & $\mathsf{TCCCCTAG} \rightleftharpoons \mathsf{CTAGGGGA}$ & 1.58451\\
3.03333 & $\mathsf{GGGCTAGG} \rightleftharpoons \mathsf{CCTAGCCC}$ & 2.15556\\
3.00000 & $\mathsf{CCTCTAGA} \rightleftharpoons \mathsf{TCTAGAGG}$ & 1.58730\\ \hline
\multicolumn{3}{c}{$f_{\omega} < \widetilde{f}_{\omega}$}\\
0.19367 & $\mathsf{CCCACGTG} \rightleftharpoons \mathsf{CACGTGGG}$ & 0.57579\\
0.20856 & $\mathsf{AACTAGTC} \rightleftharpoons \mathsf{GACTAGTT}$ & 0.46800\\
0.22489 & $\mathsf{TCGGCCTA} \rightleftharpoons \mathsf{TAGGCCGA}$ & 0.32357\\
0.23333 & $\mathsf{AACTAGAC} \rightleftharpoons \mathsf{GTCTAGTT}$ & 0.39474\\
0.23948 & $\mathsf{TAGGACTA} \rightleftharpoons \mathsf{TAGTCCTA}$ & 0.29444\\
0.25848 & $\mathsf{AAGGCCTA} \rightleftharpoons \mathsf{TAGGCCTT}$ & 0.36585\\
0.26427 & $\mathsf{GGGCCCAG} \rightleftharpoons \mathsf{CTGGGCCC}$ & 0.63974\\
0.28393 & $\mathsf{TCTTCCTA} \rightleftharpoons \mathsf{TAGGAAGA}$ & 0.58835\\
0.28571 & $\mathsf{TCTAGAGC} \rightleftharpoons \mathsf{GCTCTAGA}$ & 0.66667\\ \hline
\multicolumn{3}{c}{inversions}\\
2.53521 & $\mathsf{CTAGCCGG} \rightleftharpoons \mathsf{CCGGCTAG}$ & 0.60876\\
2.48507 & $\mathsf{ACTACTAG} \rightleftharpoons \mathsf{CTAGTAGT}$ & 0.35664\\
2.48474 & $\mathsf{GCGCCTAG} \rightleftharpoons \mathsf{CTAGGCGC}$ & 0.43735\\
2.36000 & $\mathsf{CAGTCTAG} \rightleftharpoons \mathsf{CTAGACTG}$ & 0.65000\\
2.28571 & $\mathsf{CCCTAGGT} \rightleftharpoons \mathsf{ACCTAGGG}$ & 0.57143\\
2.10000 & $\mathsf{AGGGGTCC} \rightleftharpoons \mathsf{GGACCCCT}$ & 0.28837\\
2.08669 & $\mathsf{CTCACTAG} \rightleftharpoons \mathsf{CTAGTGAG}$ & 0.45819\\
2.03636 & $\mathsf{GCTAGGGT} \rightleftharpoons \mathsf{ACCCTAGC}$ & 0.61667\\
\end{tabular}
\end{ruledtabular}
\end{table}
It should be said, that two words $\mathsf{TCCTAGAC}$ and $\mathsf{GACCCTAG}$ exhibiting
the information value $p=5.75000$ and $p=5.57333$, respectively, have no counterword
among the information valuable ones.

To validate the data presented in Tables~\ref{T1}, \ref{eck12-3}, \ref{eck12-4},
\ref{T-eck12-7} and \ref{T-eck12-8}, we show the parameters of the relevant distributions
of the words of the length $3 \leq q \leq 8$ observed for the {\sl E.coli}~K-12 complete
genome. Table~\ref{Ec-par} shows the following parameters of the words distribution over
their information value: $p^M$ ($p_m$, respectively) is the largest information value
(the lowest one, respectively) observed within a distribution, $\overline{p}$ is a mean
information value of the dictionary, $\sigma$ is the standard deviation of the
distribution, finally, $\mu$ and $\epsilon$ are the skewness and kurtosis of the
distribution, respectively. It is evident, that the complementary palindromes shown above
occupy indeed the polar positions at the distributions.

A study of longer information valuable words is possible, as well, while we shall not
consider them. The growth of the length of words yields a decay of a set of information
valuable words \cite{s1-1,s1-2}. For any nucleotide sequence, there exists the specific
thickness of a dictionary $W_q$ (denoted as $d^{\ast}$), that results in a zero value of
information capacity of the dictionary, and, in turn, in a total absence of a word with
information value $p \neq 1$.

\section{Discussion}\label{disc}
The pattern revealed in nucleotide sequences and presented here is absolutely new,
universal, and intriguing. Some issues concerning this new symmetry pattern are obvious,
some are not. Evidently, the pattern could be observed over the sequences with an
alphabet with even number of symbols. Indeed, the pattern could not be met in sequences
from an odd number of symbols alphabet, since there is now {\sl \'{a}~priori} preference
for a symbol, which must be excluded from the pattern formation. Of course, a student may
face such situation in some system based on symbol sequence, say, natural language with
alphabetic writing system; meanwhile, we shall not consider such cases here.
\begin{table}[ht]
\caption{Parameters of the distributions of the words over their information value for
{\sl E.coli}~K-12 complete genome, for $3 \leq q \leq 8$.}\label{Ec-par}
\begin{ruledtabular}
\begin{tabular}{c|cccccr}
{} & $q=3$ & $q=4$ & $q=5$ & $q=6$ & $q=7$ & \multicolumn{1}{c}{$q=8$}\\[2mm]
$p^M$ & 1.54765 & 1.39567 & 1.49606 & 1.54082 & 2.32432 & 5.75000\\
$p_m$ & 0.61033 & 0.25788 & 0.47804 & 0.04559 & 0.27455 & 0.07813\\
$\overline{p}$ & 0.99856 & 0.99902 & 1.00014 & 1.00067 & 1.00096 & 1.00374\\
$\sigma$ & 0.17119 & 0.14272 & 0.11177 & 0.10523 & 0.11573 & 0.17524\\
$\mu$ & 0.65774 & -0.54509 & -0.24212 & -0.72572 & 0.62362 & 2.48201\\
$\epsilon$ & 2.08163 & 2.55142 & 1.83826 & 8.14549 & 6.31467 & 38.04501\\
\end{tabular}
\end{ruledtabular}
\end{table}

Next, the most intriguing thing here is, that the pattern is observed at the dictionaries
developed over a single strand of DNA sequence. Since the discovery of the double helix
of DNA by J.Watson and F.Crick, the complementary rule becomes a common place and is
introduced into the secondary school programs. Here we observe the pattern through the
analysis of a single strand developed frequency dictionary, and such strand ``knows
nothing'' about the existence of the opposite one.

Surely, the observed pattern of a complementary palindromic symmetry is peculiar for
rather complicated and non-random sequences. Here a question arises towards a primacy of
the patterns under consideration: whether a double helix is of a primary nature, or the
statistically based complementarity of nucleotides. Nothing is known exactly towards this
point. Evidently, the comprehensive answer on this problem is somewhere in between. It is
quite natural to expect, that these two patterns (i.e., doubling of DNA with
complementarity between the nucleotides, and palindrome symmetry) result from a
co-evolution of a system of inheritance storage, transfer and processing.

Preliminary, we have studied more, than thirty genomes of bacteria, and seven eukaryotic
genomes. There was no entity among them, that did not manifest the complementary
palindromic symmetry. This phenomenon is universal, and the exclusions (if any) are of
the greatest scientific interest. Some information valuable words fall out of the
phenomenon, as the length $q$ grows up. One may say, that the stability of the observed
symmetry decays, as the thickness $q$ of a dictionary grows up. Meanwhile, the phenomenon
seems to be very stable, for the dictionaries of the thickness $q=3$ and $q=4$. Such
stability is peculiar for rather long genomes, at least. Indeed, the data concerning the
distribution of information valuable triplets observed for a family of phage and viral
genomes \cite{ms98} show the occurrence of the pattern, while the exclusions are more
frequent.

The composition of the counterwords composing complementary palindromes is of great
interest, itself. Various genomes exhibit different words within the couples. Meanwhile,
some words (especially, rather short ones, say, of the length $q=3$ and $q=4$) occur
quite often, in various entities. A comparative study of such words may contribute
significantly an investigation of evolutionary relations between the nucleotide sequences
and their bearers.

Interspecies variation in complementary palindromic triplets is greater, in comparison to
that on observed within a genome. A growth of the dictionary thickness $q$ yields an
expansion of a variety of complementary palindromes observed in different chromosomes of
the same genome. The greater diversity of the sets of complementary palindromic triplets
observed in a single chromosome genomes may follow from the taxonomy of these latter.
Indeed, a single chromosome genomes are peculiar for prokaryotic organisms, while the
multi-chromosomal ones are met in eukaryotic beings. Such great difference in
organisation level of inherited matter may manifest in various ways, including the
difference in complementary palindromic triplets (and quadruplets) composition.

The information valuable words themselves are known to be spread alongside a genome quite
non-randomly, with significant preference to some peculiar segments of a genome
\cite{kitai,dan}. A study of the dispersion of the counterwords composing complementary
palindromes may reveal more specific occurrence pattern of such strings alongside a
genome.

\begin{acknowledgments}
We would like to extend our gratitude to Prof.Alex\-ander~N.Gorban from the University of
Leicester for long-time collaboration and permanent encouraging interest to the work, and
to Dr.Tatyana~G.Popova who draw the attention to the problem of distribution of
information valuable words alongside a genome.
\end{acknowledgments}

\end{document}